\documentclass{article}

\usepackage{biblatex}
\usepackage{booktabs}
\usepackage[utf8]{inputenc}
\usepackage[T1]{fontenc} % T2A for cyrillics
\usepackage[fleqn]{amsmath}
\usepackage{graphicx}
\usepackage{siunitx}
\usepackage{enumitem}
\usepackage[autostyle=true]{csquotes}
\usepackage{xurl}
\usepackage{lipsum}
\usepackage[most]{tcolorbox}
\usepackage{tabularx}
\usepackage{multirow}
\usepackage{subcaption}
\usepackage{rotating}
\usepackage{amsthm}
\usepackage{sansmath}
\usepackage{xspace}
\usepackage[acronym]{glossaries}
\usepackage{cleveref}

\newacronym[description={Fourth Generation of Cellular Network Technology}]{4G}{4G}{fourth generation of cellular network technology}
\newacronym[description={Fifth-Generation Technology Standard for Cellular Networks}]{5G}{5G}{fifth-generation technology standard for cellular networks}
\newacronym[description={Sixth-Generation Technology Standard for Cellular Networks}]{6G}{6G}{sixth-generation technology standard for cellular networks}
\newacronym[description={Access and Mobility Management Function}]{AMF}{AMF}{access and mobility management function}
\newacronym[description={Application Programming Interface}]{API}{API}{application programming interface}
\newacronym[description={Autonomous System}]{AS}{AS}{autonomous system}
\newacronym[description={Affinity for Technology Interaction}]{ATI}{ATI}{affinity for technology interaction}
\newacronym[description={Authentication Server Function}]{AUSF}{AUSF}{authentication server function}
\newacronym[description={Content Delivery Network}]{CDN}{CDN}{content delivery network}
\newacronym[description={Commercial Internet Exchange}]{CIX}{CIX}{commercial Internet exchange}
\newacronym[description={Domain Name System}]{DNS}{DNS}{domain name system}
\newacronym[description={Digital Object Identifier}]{DOI}{DOI}{digital object identifier}
\newacronym[description={Disruption-Tolerant Network}]{DTN}{DTN}{disruption-tolerant network}
\newacronym[description={Edge Application Server Discovery Function}]{EASDF}{EASDF}{edge application server discovery function}
\newacronym[description={Integrated Access and Backhaul}]{IAB}{IAB}{integrated access and backhaul}
\newacronym[description={Connectivity Monitoring Function}]{CMF}{CMF}{connectivity monitoring function}
\newacronym[description={Journal Impact Factor}]{JIF}{JIF}{journal impact factor}
\newacronym[description={Information and Communication Technology},longplural={information and communication technologies}]{ICT}{ICT}{information and communication technology}
\newacronym[description={IP Multimedia Subsystem}]{IMS}{IMS}{IP multimedia subsystem}
\newacronym[description={Isolated Operation for Public Safety}]{IOPS}{IOPS}{isolated operation for public safety}
\newacronym[description={Inter-Quartile Range}]{IQR}{IQR}{inter-quartile range}
\newacronym[description={Instructed Response Item}]{IRI}{IRI}{instructed response item}
\newacronym[description={Internet Service Provider}]{ISP}{ISP}{Internet service provider}
\newacronym[description={Internet Exchange Point}]{IXP}{IXP}{Internet exchange point}
\newacronym[description={Kernel Density Estimation}]{KDE}{KDE}{kernel density estimation}
\newacronym[description={Multi-Access Edge Computing}]{MEC}{MEC}{multi-access edge computing}
\newacronym[description={Mobile Network Operator}]{MNO}{MNO}{mobile network operator}
\newacronym[description={Network Exposure Function}]{NEF}{NEF}{network exposure function}
\newacronym[description={Network Repository Function}]{NRF}{NRF}{network repository function}
\newacronym[description={Non-Terrestrial Network}]{NTN}{NTN}{non-terrestrial network}
\newacronym[description={Policy Control Function}]{PCF}{PCF}{policy control function}
\newacronym[description={Point of Presence},longplural={points of presence}]{PoP}{PoP}{point of presence}
\newacronym[description={Radio Access Network}]{RAN}{RAN}{radio access network}
\newacronym[description={Session Management Function}]{SMF}{SMF}{session management function}
\newacronym[description={Subscriber Identity Module}]{SIM}{SIM}{subscriber identity module}
\newacronym[description={Self-Reported Single Item}]{SRSI}{SRSI}{self-reported single item}
\newacronym[description={State Synchronization Function}]{SSF}{SSF}{state synchronization function}
\newacronym[description={Uncrewed Aerial Vehicle}]{UAV}{UAV}{uncrewed aerial vehicle}
\newacronym[description={Unified Data Repository}]{UDR}{UDR}{unified data repository}
\newacronym[description={Unified Data Management}]{UDM}{UDM}{unified data management}
\newacronym[description={User Plane Function}]{UPF}{UPF}{user plane function}

\newacronym{3GPP}{3GPP}{3rd Generation Partnership Program}
\newacronym{5G-A}{5G-A}{5G-Advanced}
\newacronym{5G-NR}{5G-NR}{5G New Radio}
\newacronym{ACM}{ACM}{Association for Computing Machinery}
\newacronym{BGP}{BGP}{Border Gateway Protocol}
\newacronym{CSCW}{CSCW}{Computer-Supported Collaborative Work}
\newacronym{DoS}{DoS}{Denial-of-Service}
\newacronym{GPS}{GPS}{Global Positioning System}
\newacronym{HCI}{HCI}{Human-Computer Interaction}
\newacronym{ICMP}{ICMP}{Internet Control Message Protocol}
\newacronym{ICORE}{ICORE}{International CORE Conference Rankings}
\newacronym{iGDB}{iGDB}{Internet Geographic Database}
\newacronym{IoT}{IoT}{Internet of Things}
\newacronym{IP}{IP}{Internet Protocol}
\newacronym{ISO}{ISO}{International Organization for Standardization}
\newacronym{ITU-T}{ITU-T}{International Telecommunication Union Telecommunication Standardization Sector}
\newacronym{NIST}{NIST}{National Institute of Standards and Technology}
\newacronym{O-RAN}{O-RAN}{Open Radio Access Network}
\newacronym{RRC}{RRC}{Radio Resource Control}
\newacronym{VoIP}{VoIP}{Voice over Internet Protocol}
\newacronym{WWW}{WWW}{World Wide Web}

\usepackage{framed} % for shaded env's
\usepackage{stix} % for the black hexagon symbol

\newenvironment{emphbox}
{%
  \begin{shaded}
  \noindent
}
{%
  \end{shaded}
}

\definecolor{mybeige}{HTML}{DDC1B1}
\definecolor{myblue}{HTML}{5E62B8}
\definecolor{mypurple}{HTML}{C5C6E7}
\definecolor{mygray}{HTML}{878697}
\definecolor{mylightbeige}{HTML}{F5DCC9}
\definecolor{mylightpurple}{HTML}{E8E6FE}
\colorlet{shadecolor}{mybeige!30}
\colorlet{altrowcolor}{myblue!30}

\newcolumntype{R}{>{\raggedleft\arraybackslash}X}
\newcolumntype{L}[1]{>{\raggedright\arraybackslash}p{#1}}
\newcolumntype{C}[1]{>{\centering\arraybackslash}p{#1}}

\newcommand{\tfmark}[1]{\textsuperscript{#1}}
\newcommand{\tfnote}[2]{\tfmark{#1} #2}
\newcommand{\tfnotes}[3]{\multicolumn{#1}{p{#2}}{\tiny{#3}} \\ \tiny{} \\}

\DeclareSIUnit{\year}{years}

\newcommand{\ctxcircle}{\emx{\mdsmblkcircle}}
\newcommand{\ctxsquare}{\emx{\mdsmblksquare}}

\newcommand{\captionstrongemph}[1]{{\textbf{#1}}}
\newcommand{\captionheadline}[1]{\captionstrongemph{#1.}}

\newcommand{\boxstrongemph}[1]{{\textbf{#1}}}

\newcommand{\toolname}{DRACO\xspace}
\newcommand{\eg}{e.\,g.\xspace}
\newcommand{\ie}{i.\,e.\xspace}

\newcommand{\emx}[1]{\ensuremath{#1}\xspace}
\newcommand{\actor}{\emx{y}}
\newcommand{\actors}{\emx{\mathcal{Y}}}
\newcommand{\app}{\emx{a}}
\newcommand{\apps}{\emx{\mathcal{A}}}
\newcommand{\connection}{\emx{c}}
\newcommand{\connections}[2]{\emx{\mathcal{C}(#1,#2)}}
\newcommand{\developer}{\emx{d}}
\newcommand{\developers}{\emx{\mathcal{D}}}
\newcommand{\link}{\emx{e}}
\newcommand{\links}{\emx{\mathcal{E}}}
\newcommand{\graph}{\emx{\mathcal{G}}}
\newcommand{\datacenter}{\emx{h}}
\newcommand{\datacenters}{\emx{\mathcal{H}}}
\newcommand{\user}{\emx{u}}
\newcommand{\users}{\emx{\mathcal{U}}}
\newcommand{\operator}{\emx{o}}
\newcommand{\operators}{\emx{\mathcal{O}}}
\newcommand{\provider}{\emx{p}}
\newcommand{\providers}{\emx{\mathcal{P}}}
\newcommand{\route}{\emx{r}}
\newcommand{\routes}[2]{\emx{\mathcal{R}(#1,#2)}}
\newcommand{\routesalone}{\emx{\mathcal{R}}}
\newcommand{\server}{\emx{s}}
\newcommand{\servers}{\emx{\mathcal{S}}}
\newcommand{\system}{\emx{z}}
\newcommand{\systems}{\emx{\mathcal{Z}}}
\newcommand{\timeslot}{\emx{t}}
\newcommand{\timeslots}{\emx{\mathcal{T}}}
\newcommand{\area}{\emx{v}}
\newcommand{\areas}{\emx{\mathcal{V}}}
\newcommand{\challenge}{\emx{x}}
\newcommand{\challenges}{\emx{\mathcal{X}}}
\newcommand{\availability}{\emx{\alpha}}
\newcommand{\robustness}{\emx{\omega}}
\newcommand{\utility}{\emx{\sigma}}
\newcommand{\resilience}{\emx{\rho}}

\newtheoremstyle{researchquestionstyle}
  {0pt}                                      % Space above
  {0pt}                                      % Space below
  {\normalfont}                              % Body font
  {0pt}                                      % Indent
  {\textbf}   % Heading font
  {.}                                        % Punctuation
  {0.5em}                                    % Space after heading
  {}                                         % Default heading format
\theoremstyle{researchquestionstyle}
\newtheorem{rqtheorem}{Research Question}
\crefname{rqtheorem}{RQ}{RQs}
\Crefname{rqtheorem}{RQ}{RQs}

\newlist{outlineitems}{itemize}{1}
\setlist[outlineitems]{
  label=\textcolor{red}{$\bullet$},
  leftmargin=*,
  before=\color{red},
}

\title{Extended Version: Measuring the End-to-End Resilience of Application Deployments in Real-World Communication Networks with \toolname}
\author{%
    Leon Janzen\textsuperscript{*}, Matthias Hollick\textsuperscript{*,$\dagger$} \\
    \textit{\textsuperscript{*}Secure Mobile Networking Lab} \\
    Technical University of Darmstadt \\
    Darmstadt, Germany \\
    \{ljanzen,mhollick\}@seemoo.tu-darmstadt.de \\
    \textit{\textsuperscript{$\dagger$}Secure Mobile Systems Group} \\
    IMDEA Networks Institute \\
    Madrid, Spain \\
    matthias.hollick@networks.imdea.org 
}

\begin{document}

\maketitle

\begin{abstract}
Billions of users take Internet connectivity for granted and use it daily to access applications deployed anywhere around the globe. What users perceive as a seamless connection between their end device and the application server involves a complex system of systems operated by various actors. Failures in any of these systems disrupt user-to-application connectivity if the communication network is not resilient. Since communication networks are a critical infrastructure as part of \gls{ICT}, real-world operators rarely disclose information about them. This makes it difficult to measure the end-to-end resilience of deployments, because it requires guessing topologies and the availability of the involved systems.

This paper presents DRACO, a framework for modeling communication networks and measuring end-to-end resilience of application deployments. It enables configuring application deployments across arbitrary communication networks and evaluating their availability, robustness, and resilience. We demonstrate the functional scope of DRACO by modeling nationwide application deployments in Germany and France, highlighting its compatibility with a wide variety of public data sources and synthetic data.
\end{abstract}

% ##################################################################################################################################
% ##################################################################################################################################
% ##################################################################################################################################
% ##################################################################################################################################
% ##################################################################################################################################

\section{Introduction}
The Internet is a global communication network that enables users to connect to application servers hosted as far as at the other end of the planet within milliseconds.
When all systems involved in establishing Internet connections are available, users perceive Internet connectivity as seamless.
However, that relies on the availability of a complex system of systems operated by various actors, \eg, the data center hosting the available application server and a sequence of hops from the user to the application server through the global network of terrestrial and submarine Internet transport cables~\cite{franken_2024_subsea,knight_2011_internet}.
Failures in any of these systems disrupt user-to-application connectivity if the communication network is not resilient.
In line with the dimensions of the Internet, there are countless definitions for the resilience of individual aspects of communication networks, ranging from the fault tolerance of electronic components to the resilience of cellular systems~\cite{lj-resiliencebydesign}.
Najarian and Lim~\cite{najarian_2019_design} assess the validity of resilience metrics in a case study of power systems that uses an optimization simulator and an analysis of variance design.
In contrast, we define and measure the end-to-end resilience of application deployments in communication networks, using network models and public data.
The aspect we examine in this work is the end-to-end connectivity from users to applications in communication networks.
Our first research question aims to define and measure the resilience of application deployments:

\begin{emphbox}
\textbf{RQ1.}
How can the end-to-end resilience of application deployments be defined and measured?
\end{emphbox}

The deployment strategy of an application affects the number of locations where it is hosted.
End-to-end connections from users to an application can terminate at any of these server locations, and modern \emph{high-availability} deployments feature server replicas in different regions to improve reliability through cloud and edge computing.
\Cref{sec: pub-f system model} establishes a system model for communication networks with a focus on the user-to-application connectivity.
We introduce definitions for the connectivity, availability, utility, robustness, and resilience of application deployments.
Then, we propose a method for measuring an app's end-to-end resilience based on these metrics.
Modeling realistic application deployments in communication networks is hard because public operator-verified topology and usage information are scarce, as they conflict with the safety and business interests of the involved actors.
Data on the physical locations of core Internet infrastructure is generally not public because most parts of the Internet are considered critical infrastructure, making it undesirable to disclose the exact locations of Internet sites and cables.
% On another note, most application operators keep their application deployment strategies private to deter targeted attacks.
This leaves researchers to assume and simulate information that only real-world operators can verify, making it difficult for researchers to model realistic application deployments.
We address this problem with our second research question:

\begin{emphbox}
\textbf{RQ2.}
How can application deployments in communication networks be modeled realistically with public data?
\end{emphbox}

In \Cref{sec: pub-f tool}, we present \toolname, a framework for modeling communication networks and measuring the end-to-end resilience of application deployments within them.
\toolname enables the research community to configure application deployments in communication networks using public data sources or their own data.
It facilitates the evaluation of our end-to-end resilience metric and is easily extensible to incorporate further metrics that measure other aspects of user-to-application connectivity in communication networks.
We conduct two case studies, investigating the resilience of mobile messengers in France (\Cref{sec: pub-f eval france}) and the resilience of cellular connectivity in Germany (\Cref{sec: pub-f eval germany}).
These case studies demonstrate how researchers can use \toolname with a combination of public data and synthetic data to model communication networks and evaluate application deployments within them.
% In summary, this paper offers the following contributions:
% \begin{itemize}
%     \item We provide definitions for the end-to-end availability, robustness, and resilience of application deployments in communication networks (\Cref{sec: pub-f system model}).
%     \item We present \toolname, a framework for measuring these resilience-related properties in flexibly configurable communication networks (\Cref{sec: pub-f tool}).
%     \item We demonstrate how researchers can integrate public data sources and synthetic data into \toolname (\Cref{sec: pub-f eval france,sec: pub-f eval germany}).
% \end{itemize}
\emph{We will publish the source code of \toolname and the data to reproduce our findings upon the acceptance of this paper.}

% ##################################################################################################################################
% ##################################################################################################################################
% ##################################################################################################################################
% ##################################################################################################################################
% ##################################################################################################################################

\section{System Model}
\label{sec: pub-f system model}

% This section introduces definitions to measure the resilience of application deployments in communication networks.
Communication networks can be regarded from different perspectives and modelled in many ways~\cite{bolton_1994_firm,frank_1971_computer,gomory_1964_synthesis,a-visionpaper}.
Our own interpretation best aligns with that of Janzen et al.~\cite{a-visionpaper}, which we adapt to the context of our scenario:

\begin{emphbox}
A \boxstrongemph{communication network} is a multi-actor system of systems that enables \emph{connectivity} between users and applications.
Connectivity involves numerous systems under the control of multiple actors who have to cooperate to provide communication as an end-to-end service.
A \boxstrongemph{resilient} communication network is prepared to face \emph{challenges} that affect systems, withstand them, and prevent most challenges from causing \emph{utility} degradation.
\end{emphbox}

\begin{figure}
    \includegraphics[width=\columnwidth]{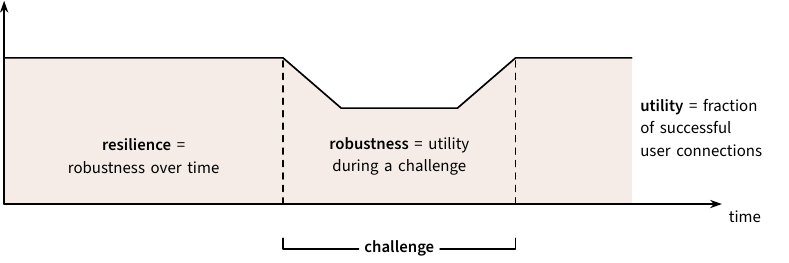}
    \caption[Utility, Robustness, and Resilience during Challenges]{%
        \captionheadline{Utility, Robustness, and Resilience during Challenges}
    }
    \label{fig: pub-f resilience theory}
\end{figure}

\Cref{fig: pub-f resilience theory} visualizes our interpretation of resilience and its relation to utility, challenges, and robustness.
The \emph{service} provided by communication networks is the end-to-end connectivity between users and application servers.
The \emph{utility} of that service is the quota of successful user-to-application connections, \ie, the fraction of users that successfully access their desired application through the communication network.
The following definitions of availability, robustness, and resilience align with  \gls{ISO}~\cite{iso-24765-2010}, the \gls{NIST}~\cite{nist_2022_engineering}, and the \gls{ITU-T}~\cite{itut_1994_qos}.
End-to-end connectivity involves many systems, and the \emph{availability} of a system is the (theoretical) probability that a system provides its intended service at any given time.
The \emph{robustness} of a system translates its availability into practice, \ie, it determines the level to which a system provides its intended purpose at a specific time under specific circumstances.
These circumstances can include \emph{challenges} that cause failures in the communication network, potentially disrupting connectivity between users and applications.
The \emph{resilience} accounts for a system's preparedness for arbitrary challenges, measuring its robustness over time.
% This section introduces a system model for deployments in communication networks that jointly models aspects of communication networks and application deployments to account for the end-to-end nature of user-to-application connections.
% We present definitions for resilience-related properties and define a metric that measures the end-to-end resilience of applications.

\subsection{Notation}
\label{ssec: pub-f system model notation}

\begin{table}
  \begin{tabularx}{\linewidth}{llX}
    \toprule
    &
    \textbf{Notation} &
    \textbf{Description} \\
    \midrule
    \multirow{4}*{Topology} &
    $\area\in\areas$ & Areas in the communication network \\
    & $\link\in\links$ & Links connecting the areas \\
    & $\graph=(\areas,\links)$ & Graph representing the network topology \\
    & $\route\in\routesalone$ & Routes between the areas \\
    \midrule
    \multirow{3}*{Systems} &
    $\server\in\servers$ & Application servers that users connect to \\
    & $\datacenter\in\datacenters$ & data centers hosting application servers \\
    & $\system\in\systems=\servers\cup\datacenters\cup\links$ & Systems of the communication network \\
    \midrule
    \multirow{4}*{Actors} &
    $\developer\in\developers$ & Application developers operating applications \\
    & $\operator\in\operators$ & Network operators operating connections \\
    & $\provider\in\providers$ & Cloud/Edge providers hosting data centers \\
    & $\actor\in\actors=\developers\cup\operators\cup\providers$ & The set of all actors \\
    \midrule
    \multirow{4}*{Traffic} &
    $\app\in\apps$ & Applications \\
    & $\user\in\users$ & Users \\
    & $\users_\area, \users_\app, \users_{\area, \app}$ & Users in an area, of an app, and both \\
    & $\connection\in\connections{\area}{\app}$ & Connections of an area to an application \\
    \midrule
    \multirow{2}*{Chall} & 
    $\challenge\in\challenges$ & Challenges to the availability of the application \\
    & $\timeslot\in\timeslots$ & Timeslots in which challenges occur \\
    \midrule
    \multirow{4}*{Metrics} &
    $\availability:\systems\cup\apps\to[0,1]$ & Availability \\
    & $\utility:\systems\cup\apps\to[0,1]$ & Utility \\
    & $\robustness:\systems\cup\apps\to[0,1]$ & Robustness \\
    & $\resilience:\systems\cup\apps\to[0,1]$ & Resilience \\
    \bottomrule
  \end{tabularx}
  \caption[System Model Notation]{%
      \captionheadline{System Model}
      Notation of variables and functions.
  }
  \label{tab: pub-f notation}
\end{table}

% For better readability, we only introduce selected notation in the text.
\Cref{tab: pub-f notation} shows the entire system model notation relevant for our work.
%\subsubsection{Communication Networks.}
The graph $\graph=(\areas,\links)$ describes the topology of the communication network.
Areas $\area\in\areas$ have geographic locations that we use to calculate link lengths and to visualize the communication network.
Each link $\link\in\links$ connects two areas.
Data centers $\datacenter\in\datacenters$ are located in an area, and servers $\server\in\servers$ are hosted in data centers.
Communication networks are complex systems of systems operated by multiple actors, all of whom play a role in connecting users to applications.
Application developers operate and deploy applications within the communication network.
Users access applications via the communication network.
Cloud and edge providers operate data centers, where application servers are hosted.
Network operators maintain links to connect users across areas.
Each application is deployed across a set of application servers hosted in a subset of the available data centers.
In each area, users access applications.
% There are users in each area who access applications.

\subsection{Application Deployment Resilience}
\label{ssec: pub-f system model graph metrics}

%\subsubsection{Connectivity.}
% Many systems in the communication network are involved in connecting application users to an application server, and there are often many potential routes through the communication network that connect a user to one of the application servers.
Let \routes{\area_s}{\area_t} denote all inter-area routes that connect source area $\area_s$ and target area $\area_t$.
We define routes as loop-free lists of links, where the first link starts from $\area_s$ and the last link ends in $\area_t$.
Next, we introduce end-to-end connectivity, which describes the potential routes that user traffic can take to the app.

\begin{emphbox}
The \boxstrongemph{end-to-end connectivity} of the users accessing application \app from area \area is the set of routes that connect them to any area where an application server $\server_\app$ is hosted:
\begin{equation}
\label{eq: pub-f connectivity}
    \connections{\area}{\app} = \bigcup_{\area_\app\in \areas_\app} \routes{\area}{\area_\app},
\end{equation}
where $\areas_\app\subseteq\areas$ models the areas where a server of application \app is hosted.
\end{emphbox}

%\subsubsection{Utility.}
A system's utility $\utility(\system)$ is the degree to which the system \system accomplishes its designated functions~\cite{iso-24765-2010}.
For instance, the function of an inter-area link is to connect users and applications across two areas, the function of a data center is to provide access to application servers, and the function of an application server is to provide users with access to an app.
Let $\utility(\area, \app)$ model whether the users in area \area can access application \app.
This allows us to model the end-to-end utility of applications based on the fraction of users who can access them.

\begin{emphbox}
    The \boxstrongemph{end-to-end utility} of an application \app is the quota of successful user connections:
    \begin{equation}
    \label{eq: pub-f utility}
        \utility(\app) = \frac{1}{\users_\app} \cdot \sum_{\area\in\areas} \users_{\area,\app}\cdot\utility(\area,\app),
    \end{equation}
    where $\utility(\area, \app)$ models whether users in area \area can access application \app.
\end{emphbox}

%\subsubsection{Availability.}
A system's availability is the probability that it accomplishes its designated functions at any given time~\cite{iso-24765-2010}.
The availability of link~\link is the chance that it connects users across areas at an arbitrary point in time.
Let $\availability(\system)$ denote the availability of systems $\system\in\systems$ in the communication network, \eg, $\availability(\link)$ for links, $\availability(\datacenter)$ for data center, and $\availability(\server)$ for application servers.
Let $\availability(\route)=\prod_{\link\in\route} \availability(\link)$ denote the availability of route \route as the probability that all links of the route are available.
We model the end-to-end availability of an application \app for the users in area \area as
\begin{equation*}
\label{eq: pub-f availability-per-ares}
    \availability(\area,\app) = 1 - \prod_{\route\in\connections{\area}{\app}}
    \big(
        1 - \availability(\route)\cdot\availability(\datacenter)\cdot\availability(\server_\app)
    \big)
\end{equation*}
where \datacenter is the data center at the end of route \route, and $\server_\app$ the server of application \app that is hosted in data center \datacenter.
The calculation of $\availability(\area,\app)$ is based on the complement probability that there is no available route to an area with an available application server.
We generalize this definition from individual areas to the entire network:

\begin{emphbox}    
The \boxstrongemph{end-to-end availability} of an application \app describes the probability that any user can access the application from any area at any point in time:
\begin{equation}
\label{eq: pub-f availability}
    \availability(\app) = \frac{1}{\users_\app} \cdot \sum_{\area\in\areas} \users_{\area,\app} \cdot \availability(\area,\app)
\end{equation}
\end{emphbox}

%\subsubsection{Challenges.}
A \emph{challenge} causes systems to fail for a number of timeslots~\cite{iso-24765-2010}.
% This introduces the concept of time into our system model.
In every timeslot, users try to access applications, and the end-to-end utility of the application deployment captures the fraction of successful user-to-application connections.
The end-to-end robustness of application deployments captures their utility during a timeslot with an active challenge.
% After that, we show how to measure resilience by considering utility/robustness over time.
%\subsubsection{Robustness.}
We consider \emph{robustness} to be the degree to which a system or component can function correctly in the presence of challenges~\cite{iso-24765-2010}.
We model the robustness of system $\system\in\systems$ to challenge $\challenge\in\challenges$ as
\begin{equation*}
\robustness(\system,\challenge)=
\begin{cases}
    1 & \text{if \system provides its service during \challenge} \\
    0 & \text{if \system does not provide its service during \challenge}. \\
\end{cases}
\end{equation*}

We model the robustness of application \app to challenge \challenge for the users in area \area as
\begin{equation*}
    \robustness(\area,\app,\challenge) = 1 - \prod_{\route\in\connections{\area}{\app}}
    \big(
        1 - \robustness(\route,\challenge)\cdot\robustness(\datacenter,\challenge)\cdot\robustness(\server_\app,\challenge)
    \big) 
\end{equation*}
where \datacenter is the data center at the end of route \route, the server $\server_\app$ is hosted in data center \datacenter, and $\robustness(\area,\app,\challenge)$ models whether users in area \area can access application \app in the face of challenge \challenge.
Finally, we derive the robustness of applications \app to challenge \challenge:

\begin{emphbox}
The \boxstrongemph{end-to-end robustness} of an application \app to challenge \challenge describes the quota of users in the communication network who wish to access the application \app and can successfully do so in the face of challenge \challenge:
\begin{equation}
\label{eq: pub-f robustness}
    \robustness(\app,\challenge) = \frac1{\users_\app} \cdot \sum_{\area\in\areas} \users_{\area,\app} \cdot \robustness(\area,\app,\challenge)
\end{equation}
\end{emphbox}

%\subsubsection{Resilience.}
For the notion of \emph{resilience}, we consider applications deployed in a communication network over a period of time $\timeslots = [\timeslot_0, \timeslot_1, \dots, \timeslot_{|\timeslots|-1}]$, modeled as a sequence of timeslots.
All aforementioned metrics can be measured for individual timeslots. %, \eg, the robustness $\robustness(\app,\challenge)$ measures the quota of successful user connections to application \app during a timeslot with active challenge \challenge.
In contrast, resilience measures a system over a sequence of timeslots.

\begin{emphbox}
    The \boxstrongemph{end-to-end resilience} of an application \app is the level to which it withstands arbitrary challenges \challenge and remains accessible for users during the affected timeslots:
\begin{equation}
\label{eq: pub-f resilience}
    \resilience(\app,\challenge,\timeslots) = \int_{\timeslot_0}^{\timeslot_{|\timeslots|-1}} \robustness(\app,\timeslot) \, dt,
\end{equation}
where $\robustness(\app,\timeslot)$ measures the robustness of \app to challenges \challenge in timeslot \timeslot.
\end{emphbox}

% In this section, we addressed research question \Cref{rq: e.1} that sought a way to define and measure the end-to-end resilience of application deployments:
% \begin{emphbox}
%     \strongemph{Answer to \Cref{rq: e.1}:}
%     Our end-to-end resilience metric measures the area under the end-to-end robustness graph that plots the quota of successful user-to-application connections over time.
%     This definition accounts for an application deployment's average utility over time, its robustness against arbitrary challenges, and the recovery speed after challenges.
% \end{emphbox}

% We implement the metrics presented in this section in our resilience measurement framework \toolname that we introduce next.

% ##################################################################################################################################
% ##################################################################################################################################
% ##################################################################################################################################
% ##################################################################################################################################
% ##################################################################################################################################

\section{\toolname: An End-to-End Resilience Measurement Framework}
\label{sec: pub-f tool}

We present a resilience measurement framework \toolname.\footnote{\toolname is an anagram of \enquote{\underline{c}onnectivity \underline{r}esilience \underline{o}f \underline{a}pplication \underline{d}eployments}}
It enables users to model communication networks, deploy applications within them, and measure their end-to-end availability, utility, robustness, and resilience under various challenges.

\subsection{Workflow}

\begin{figure}
     \includegraphics[width=\columnwidth]{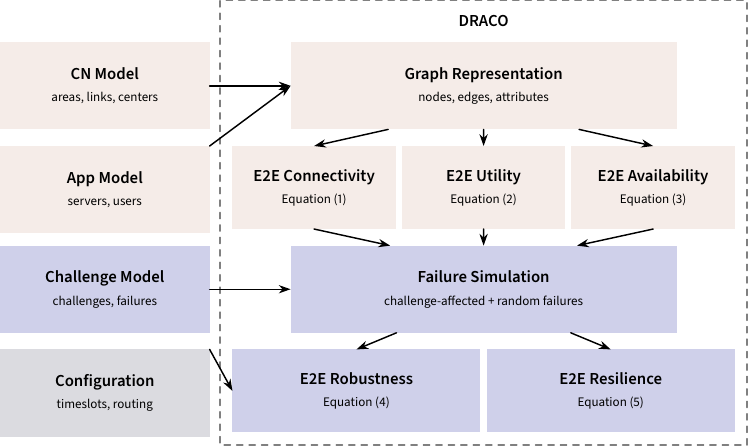}
    \caption[Evaluation Workflow]{%
        \captionheadline{Evaluation Workflow}
        Users upload a configuration with models for a communication network (CN), an application deployment, and challenges, and \toolname calculates the end-to-end metrics defined in \Cref{eq: pub-f connectivity,eq: pub-f utility,eq: pub-f availability,eq: pub-f robustness,eq: pub-f resilience}.
    }
    \label{fig: pub-f draco workflow}
\end{figure}

\Cref{fig: pub-f draco workflow} visualizes the workflow to evaluate scenarios with \toolname.
First, the user uploads files that specify the communication network, with its areas, links, and data center locations, and the application deployment, with application servers and users per area.
\toolname processes the user input and generates an intermediary graph representation of areas (\emph{nodes}) and links (\emph{edges}).
The node attributes model the data center locations, application servers, and the application users per area, as well as the availability of data centers and application servers.
The edge attributes model the links' distances and their availability.
Then, the user uploads files that specify the scenario, \ie, challenges and timeslots.
\toolname processes the user input and starts the failure simulation.
For each timeslot, active challenges are evaluated to determine the affected systems.
Systems affected by a challenge are simulated to fail in that timeslot.
For unaffected systems, \toolname conducts probabilistic testing based on the system's availability to determine whether it fails in that timeslot.
\toolname displays the measurement results, visualizations of the intermediate graph representation, and graph-related metrics with interpretation guidelines.
% The user can zoom in to gain deeper insights into the end-to-end statistics of individual areas, applications, or timeslots.
% The user can download selected measurement results.

The following two sections demonstrate \toolname's functionality and its support for a wide range of input data sources by conducting two case studies.
We evaluate the resilience of two scenarios: mobile applications in France (\Cref{sec: pub-f eval france}) and cellular connectivity in Germany (\Cref{sec: pub-f eval germany}).

% ##################################################################################################################################
% ##################################################################################################################################
% ##################################################################################################################################
% ##################################################################################################################################
% ##################################################################################################################################

\section{Case Study: Mobile Messaging in France}
\label{sec: pub-f eval france}

\begin{table}
    \begin{tabularx}{\linewidth}{lXX}
        \toprule
        &
        \textbf{Case Study 1 (France)} &
        \textbf{Case Study 2 (Germany)} \\
        \midrule
        Areas & 19 metropolitan areas\tfmark{a} & 42 cities \\
        Links\tfmark{a} & 39 physical fiber links\tfmark{b} & 123 physical fiber links\tfmark{b} \\
        \midrule
        data centers & 7 data centers\tfmark{c} & 17 data centers\tfmark{c} \\
        Applications & 6 messengers\tfmark{a} & 3 fictive \gls{5G} systems \\
        Deployments & 1 server per application & 1,4,11 servers per application \\
        \midrule
        Users & Reported application traffic\tfmark{a} & Derived from population \\
        Timeslots & One day (96 15-min slots)\tfmark{a} & One day (96 15-min slots) \\
        \bottomrule
        \tfnotes{3}{\linewidth}{%
        \tfnote{a}{NetMob~\cite{martinez_2023_netmob}}
        \tfnote{b}{iGDB~\cite{anderson_2022_igdb}}
        \tfnote{c}{AWS~\cite{aws_2026_edge}, Cloudflare~\cite{cloudflare_2026_edge}, GCP~\cite{google_2025_cloud}, Meta~\cite{meta_2026_edge}}
        }
    \end{tabularx}
    \caption[Input Data]{%
        \captionheadline{Input Data}
        Data sources used in our case studies.
    }
    \label{tab: pub-f data-sources}
\end{table}

\Cref{tab: pub-f data-sources} summarizes the data sources integrated into both case studies
% The goal of this case study is to model the Internet in France as a communication network of users accessing applications.
We choose the application category \emph{messengers} because messaging is one of the most prominent use cases for mobile applications during everyday life and in crisis situations~\cite{b-userperspective}.

\subsection{Input Data}
We use two main data sources, \emph{NetMob}~\cite{martinez_2023_netmob} and \emph{iGDB}~\cite{anderson_2022_igdb}, to model mobile application usage in France.
In 2023, Orange published the NetMob23 dataset as part of a data challenge.
It contains real-world mobile traffic data from 68 applications, collected in 20 French cities over 77 days in April and May 2019.
NetMob has a geospatial resolution of \qtyproduct{100x100}{\meter} tiles and a temporal resolution of 15-minute timeslots, and iGDB combines the physical and logical areas of the Internet, reporting real-world transport links connecting physical nodes and combine them with \gls{IP}-level and \gls{AS}-level information.
%\subsubsection{Communication Network.}
Since iGDB does not report connections from the city of Metz, we omit this area to achieve a connected communication network.
We use the remaining 19 NetMob metropolitan areas and aggregate the NetMob traffic data per city to estimate the number of application users per area.
We use iGDB to extract real-world links that exist between the NetMob cities.
We searched online for the European data center networks of major cloud and edge providers, \ie, \emph{AWS}~\cite{aws_2026_edge}, \emph{Cloudflare}~\cite{cloudflare_2026_edge}, \emph{GCP}~\cite{google_2025_cloud}, and \emph{Meta}~\cite{meta_2026_edge}, which together operate seven data centers in France.
%\subsubsection{Application Deployments.}
We select six messengers, \emph{Apple iMessage}, \emph{Facebook Messenger}, \emph{Skype}, \emph{Snapchat}, \emph{Telegram}, and \emph{WhatsApp}, from the 68 NetMob applications.
We assign each data center to the closest area and simulate the deployment of application servers of these messengers across the data centers.
Note how the available public datasets allow us to choose real-world approximations for the areas, links, data centers, and users of this example.
As a result, we have to simulate only the deployment locations of the application servers.
For instance, we simulate an Apple iMessage server in the AWS data center of Marseille, a Facebook Messenger server in the Cloudflare data center of Lyon, and a Snapchat server in the Cloudflare data center of Bordeaux.

%\subsubsection{Challenge Scenario.}
To study the robustness of the modeled application deployments, we simulate an attack on France that challenges the availability of the deployed messengers.
% The simulated hybrid attack consists of two particular challenges that occur within 24 hours:
% First, we simulate a \gls{DoS} attack that shuts down all Cloudflare data centers for three hours from 6:30 to 9:30.
At 12:00, we simulate a targeted \gls{DoS} attack on the physical Internet infrastructure around Paris, destroying all links from Paris to the south of France. %, \ie, the connections to Marseille, Lyon, Montpellier, Orleans, Bordeaux, Rennes, Dijon, Le Man, Tours, and Nantes.
We simulate a recovery of the links to Marseille, Lyon, and Dijon at 13:00, to Montpellier at 15:00, and to Bordeaux, Rennes, Nantes, Le Mans, and Tours at 19:00.

\subsection{Interpretation of the \toolname Results}

\begin{figure}
  \includegraphics[width=.5\columnwidth]{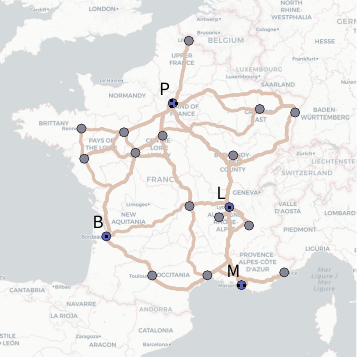}
    \caption[Case Study 1: Internet Topology of France]{%
      \captionheadline{French Internet Topology}
      A map of cities (\textcolor{mygray}{\ctxcircle}) connected by fiber links (beige lines).
      Cities marked in blue (\textcolor{myblue}{\ctxcircle}) have local data centers (\textcolor{black}{\ctxsquare}).
    }
    \label{fig: pub-f graph france}
\end{figure}

\begin{figure}
  \includegraphics[width=\columnwidth]{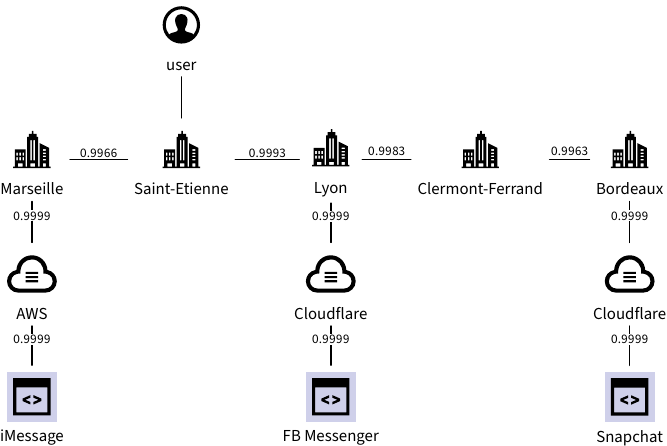}
    \caption[Case Study 1: End-to-End Availability]{%
      \captionheadline{End-to-end Availability (Case Study 1)}
      How users from Saint-Etienne access selected messengers.
    }
    \label{fig: pub-f availability france}
\end{figure}
\begin{figure}
    \includegraphics[width=\columnwidth]{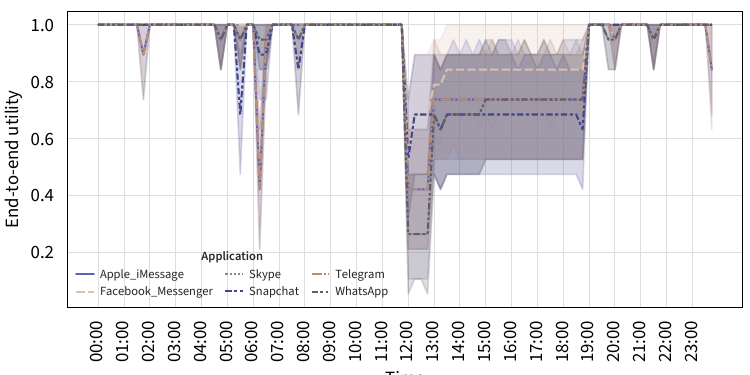}
    \caption[Case Study 1: End-to-End Resilience]{
      \captionheadline{End-to-end Resilience (Case Study 1)}
      The simulated attack on Paris leads to utility degradation visible from 12:00 to 19:00.
      The line plot shows the mean and 95\% confidence interval.
    }
    \label{fig: pub-f resilience france}
\end{figure}

\Cref{fig: pub-f graph france} visualizes the intermediate graph representation of \toolname that shows the NetMob metropolitan areas with real-world data center locations marked with red circles and the iGDB physical Internet connections between areas.
%\subsubsection{Availability.}
\Cref{fig: pub-f availability france} depicts the availability of selected messengers from the perspective of users in Saint-Étienne.
The depicted routes show the shortest paths.
\toolname calculates the end-to-end availability of each messenger from Saint-Étienne as defined in \Cref{eq: pub-f availability} with the following selected results:
The end-to-end availability of iMessage, Facebook Messenger, and Snapchat for the users in Saint-Étienne is \num{0.9966}, \num{0.9993}, and \num{0.9939}, respectively.

%\subsubsection{Utility.}
An app's utility translates the theoretical availability into practice by probabilistic probing to determine whether the systems of the communication network are functional during a timeslot.
The utility of an application drops below one when a system in the communication network is unavailable, as shown as noise in \Cref{fig: pub-f resilience france}.
\toolname determines the failure chances in the absence of challenges with probabilistic probing, considering the system's availability.
%\subsubsection{Robustness.}
An app's robustness measures its utility in the face of a challenge.
The Skype and WhatsApp servers are hosted in Paris, which explains the low robustness of these applications against the \gls{DoS} attack on Paris visible in \Cref{fig: pub-f resilience france}.
In contrast, applications with servers hosted outside of Paris, such as Snapchat, show higher robustness against the simulated attack.
%\subsubsection{Resilience.}
An app's resilience measures its utility over time, capturing its robustness against arbitrary challenges.
\toolname calculates the end-to-end resilience as defined in \Cref{eq: pub-f resilience}, with the following results:
The end-to-end resilience of the simulated deployments of Apple iMessage, Facebook Messenger, and Snapchat is \num{0.9769}, \num{0.9589}, and \num{0.9682}, respectively.

\section{Case Study: Cellular Cores in Germany}
\label{sec: pub-f eval germany}

\begin{figure}
  \includegraphics[width=.5\columnwidth]{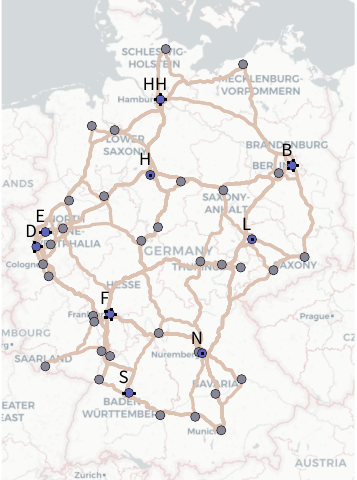}
  \caption[Case Study 2: Internet Topology of Germany]{%
      \captionheadline{German Internet Topology}
      A map of cities (\textcolor{mygray}{\ctxcircle}) connected by fiber links (beige lines).
      Cities marked in blue (\textcolor{myblue}{\ctxcircle}) have local data centers (\textcolor{black}{\ctxsquare}).
  }
  \label{fig: pub-f graph germany}
\end{figure}
\begin{figure}  
  \includegraphics[width=\columnwidth]{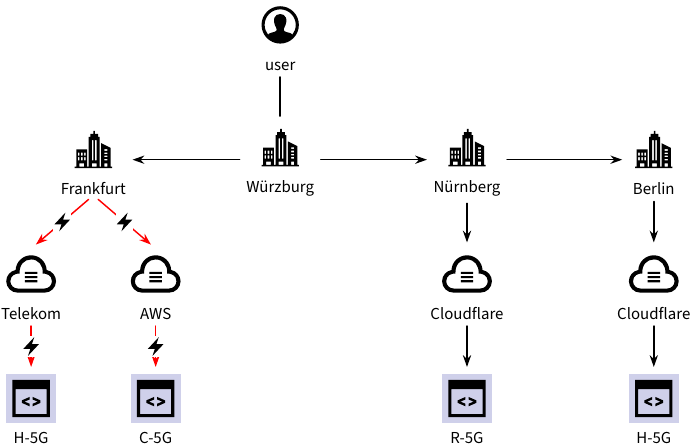}
  \caption[Case Study 2: End-to-End Robustness]{%
  \captionheadline{End-to-end Robustness (Case Study 2)}
  Simulated \gls{5G} systems with central (C-5G), high-availability (H-5G), and island-ready (R-5G) deployments.
  }
  \label{fig: pub-f robustness germany}
\end{figure}
\begin{figure}
    \includegraphics[width=\columnwidth]{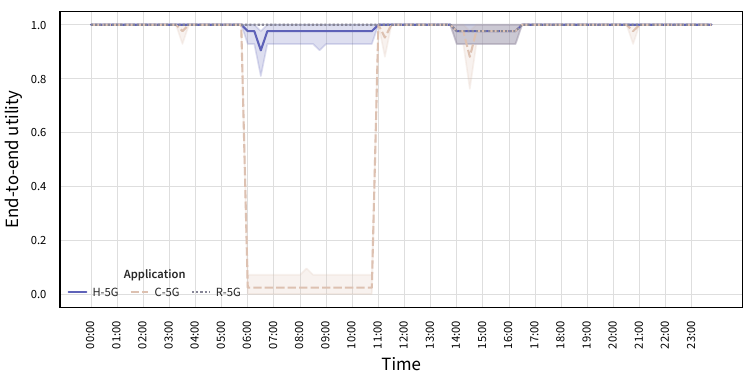}
    \caption[Case Study 2: End-to-End Resilience]{
        \captionheadline{End-to-end Resilience (Case Study 2)}
        Simulated \gls{5G} networks in Germany with simulated data center outages in Frankfurt (06:00~-~11:00) and Munich (14:00~-~16:30).
        The line plot shows the mean and 95\% confidence interval.
    }
    \label{fig: pub-f resilience germany}
\end{figure}

The goal of this case study is to evaluate the resilience of different \gls{5G} core deployment strategies.
Cellular networks heavily rely on a cellular core system that handles all control-plane tasks and connects mobile users to the Internet~\cite{3gpp-23-501}.
With the transition from \gls{4G} to \Gls{5G}, the cellular core has become cloud-native and can be hosted in arbitrary data centers.
This invites modeling \emph{\gls{5G} cores} as applications that are deployed within a country's communication network.
Since cellular networks are critical infrastructure in most countries, \glspl{MNO} do not disclose their network topologies or user traffic data.
Amini et al.~\cite{amini_2024_where} reviewed publicly available cellular datasets, but none of them are applicable to this scenario.
Therefore, we can use this case study to demonstrate how our resilience measurement framework supports synthetic and simulated data.
\Cref{tab: pub-f data-sources} summarizes the data sources integrated into this example.

\subsection{Input Data}

%\subsubsection{Communication Network.}
We use the \emph{Python} package \texttt{geonamescache} to select and model the major German cities with at least \si{100000} inhabitants as areas.
We use the iGDB project to select and model real-world links between these areas.
German \glspl{MNO} operate 17 data centers in Germany, as \glspl{ISP} increasingly offer \gls{MEC} on their own edge infrastructure.
For instance, \emph{Telekom} operates five and plans to build at least 10 further edge data centers in Germany in 2026~\cite{telekom_2025_edge}, \emph{Telefonica/O2} operates 12 data centers~\cite{o2_2026_edge}, and \emph{Vodafone} uses \emph{AWS Wavelength}~\cite{vodafone_2021_edge}.
We map these data center locations to the closest area.
%\subsubsection{Application Deployments.}
For the application servers, we model three hypothetical German cellular networks to simulate different variants of cellular core deployments:
C-5G only features a \emph{central} core hosted in the AWS data center in Frankfurt, H-5G features a \emph{high-availability} core with replicas in five data centers (Frankfurt, München, Düsseldorf, Hamburg, Berlin), and R-5G features an \emph{island-ready} \gls{5G} core with one replica in all areas with a data center (Hannover, Essen, Leverkusen, Stuttgart, Nürnberg, Leipzig, and the aforementioned locations).
According to related work, H-5G resembles today's typical \gls{5G} deployments~\cite{poignee_2025_pobtog} and R-5G implements a proposed design for resilient \gls{5G} deployments~\cite{a-visionpaper}.
We use the population of these cities as a proxy for the number of cellular users and consider two targeted attacks on German data centers:
All data centers in Frankfurt fail between 06:00 and 11:00, and all data centers in München fail between 14:00 and 16:30.
This affects the Frankfurt data centers of all \gls{5G} deployments, and those of \emph{H-5G} and \emph{R-5G} in München.

\subsection{Interpretation of the \toolname Results}

\Cref{fig: pub-f graph germany} shows the intermediate graph representation of the German Internet as a communication network of interconnected areas with data center locations marked with red circles.
%\subsubsection{Availability.}
% \Cref{fig: pub-f availability germany} depicts the end-to-end availabilities of the simulated central, high-availability, and island-ready \gls{5G} deployments.
% Of the three, \emph{R-5G} has the highest availability. % demonstrating that deploying servers across many areas increases the likelihood that users can connect to one of them.
%\subsubsection{Robustness.}
\Cref{fig: pub-f robustness germany} depicts the robustness of Würzburg during the simulated attack on Frankfurt.
The only C-5G core is hosted in the AWS data center of Frankfurt, which is not accessible from Würzburg because all Internet links to Frankfurt have failed.
The H-5G core in Frankfurt is inaccessible for the same reason, but the H-5G users in Würzburg can instead access the H-5G core in Berlin, which is accessible via Nürnberg.
The closest C-5G core to Würzburg is hosted in Nürnberg, so the C-5G users in Würzburg are unaffected by the failure of the Frankfurt C-5G core.
In contrast, the simulated attack on München has little effect on C-5G.
%\subsubsection{Resilience.}
The end-to-end resilience of the simulated \gls{5G} deployment are \num{0.7899} for \emph{C-5G}, \num{0.9164} for \emph{H-5G}, and \num{0.9692} for \emph{R-5G}.
This measures the deployments' robustness against arbitrary challenges, specifically the two simulated attacks on Frankfurt and München in this case study, visible in \Cref{fig: pub-f resilience germany}.

% \begin{emphbox}
%     \strongemph{Answer to \Cref{rq: e.2}:} Our resilience measurement framework, \toolname, allows the configuration of real-world communication networks.
%     The case studies of \Cref{sec: pub-f eval france,sec: pub-f eval germany} demonstrate \toolname's compatibility with a wide variety of public data sources.
% \end{emphbox}

% ##################################################################################################################################
% ##################################################################################################################################
% ##################################################################################################################################
% ##################################################################################################################################
% ##################################################################################################################################

\section{Discussion}
\label{sec: pub-f discussion}

% In this section, we differentiate our contributions from related work and advertise key features of \toolname that make it usable for the research community.

% \label{ssec: pub-f data availability}

The pain point in conducting security and resilience research in the realm of critical infrastructure is the lack of public operator-verified information, which forces researchers to make assumptions whose accuracy only real-world actors can verify.
We address this issue by designing \toolname to be usable with public real-world data and with synthetic data.
% This section discusses the available public data sources and the tools that can generate relevant synthetic data.
%\subsubsection{Communication Network Topologies.}

\subsection{Data Availability}

The Internet is a complex network of networks with \emph{physical} \glspl{PoP} all around the globe and links spanning land, sea, and space.
On top of that, the Internet comprises \emph{logical} networks of networks, such as \gls{IP} networks and \glspl{AS}.

\paragraph{Datasets}
% Knight et al.~\cite{knight_2011_internet} transcribed public network maps into \gls{PoP}-level.
Researchers can use iGDB to link the logical and physical layers of the Internet~\cite{anderson_2022_igdb}.
It builds on public locations and transport network routes, such as \glspl{PoP}, \glspl{IXP}, submarine cables, and their landing sites.
They interpolate connections between sites based on other critical infrastructure networks, such as roads, rails, and power lines.

\paragraph{Topology Generators}
Decades of research on Internet topologies produced a multitude of topology generators, starting with \emph{Waxman} random graphs~\cite{waxman_2002_routing} and power-law generators~\cite{aiello_2001_random}.
There are topology generators focusing on the physical Internet, such as \emph{HOT}~\cite{li_2004_first} for router-level topologies and \emph{COLD}~\cite{bowden_2014_cold} for \gls{PoP}-level topologies.
For the logical Internet, there are \gls{AS}-level topology generators, such as \emph{INET}~\cite{winick_2002_inet}, \emph{RealNET}~\cite{cheng_2008_realnet}, and Chang et al.~\cite{chang_2006_peer}.
In addition, \emph{IGEN}~\cite{quoitin_2009_igen} generates \gls{IP}-level topologies.
\emph{POBTOG}~\cite{poignee_2025_pobtog} allows researchers to generate country-wide communication network topologies based on population density.

\paragraph{Data Center Locations}
Most hyperscalers publish locations where they operate cloud data centers, \eg, AWS~\cite{aws_2026_edge}, Cloudflare~\cite{cloudflare_2026_edge}, GCP~\cite{google_2025_cloud}, and Meta~\cite{meta_2026_edge}.
In addition, an increasing number of \glspl{MNO} and \glspl{ISP} operate data centers, \eg, Telekom~\cite{telekom_2025_edge} and Telefonica/O2~\cite{o2_2026_edge} in Germany.
\emph{Together with the \toolname source code, we will publish the German and French data center locations we identified for the case studies in \Cref{sec: pub-f eval france,sec: pub-f eval germany}.}

\paragraph{Application Deployments}
The server infrastructure of most applications is not publicly known, but related work offers tools to explore the cloud usage of individual applications, \eg, \emph{CloudAnalyzer} by Henze et al.~\cite{henze_2017_cloudanalyzer} and the \emph{MATAdOR} framework by Scheitle et al.~\cite{scheitle_2016_analyzing}.
With the \emph{NetMob23} data challenge~\cite{martinez_2023_netmob}, Orange provided a dataset reporting uplink and downlink traffic for 68 mobile applications collected across 20 French cities over a period of 77 days in 2019.
However, NetMob was only shared with participants of the data challenge and is not publicly available.
The \emph{MIRAGE-APPxACT-2024} dataset~\cite{guarino_2024_mirage} features traffic traces from many Android applications, including the messengers studied in our case study of mobile messaging in France.
In addition, the \emph{MopEye} dataset~\cite{wu_2017_mopeye} features crowdsourced measurements of Android applications.

\paragraph{Challenges}
There are datasets reporting on past natural disasters, such as tropical storms~\cite{ocha_2018_storms}, earthquakes~\cite{giardini_1999_global}, and \emph{EM-DAT}~\cite{emdat_2026_international} reporting general disasters.
Clare et al.~\cite{clare_2023_climate} investigate potential challenges for Internet infrastructure caused by climate change, and Jyuoti et al.~\cite{jyothi_2021_solare} study potential solar superstorms.
Combining these with geographic models of the earth, \eg, a global relief model~\cite{etopo_2024_relief}, allows simulating natural disasters as challenges.

\begin{emphbox}
    \textbf{Answers to our research questions.}
    Our end-to-end resilience metric measures the area under the end-to-end robustness graph that plots the quota of successful user-to-application connections over time.
    This definition accounts for an application deployment's average utility over time, its robustness against arbitrary challenges, and the recovery speed after challenges (RQ1).
    Our resilience measurement framework, \toolname, allows the configuration of real-world communication networks.
    The case studies of \Cref{sec: pub-f eval france,sec: pub-f eval germany} demonstrate \toolname's compatibility with a wide variety of public data sources (RQ2).
\end{emphbox}

\subsection{Related Work}

This paper makes contributions in two realms:
We introduced a definition and metric for the end-to-end resilience of applications (RQ1) and developed a resilience measurement framework that implements it (RQ2).
Additionally, we demonstrated \toolname's functionality with two case studies.
%\subsubsection{Measuring Resilience.}

The resilience metric closest to our work is Giotsas et al.~\cite{giotsas_2025_regional} that analyzes the route resilience of the Internet from a cloud provider's perspective.
Their robustness definition is the system's ability to withstand a certain challenge with limited performance degradation, which closely resembles our own definition.
They define Internet resilience as the network's ability to maintain diverse, secure routing paths under stress.
In contrast, we define end-to-end resilience as successful user-to-app connections over time, adding a user-centric perspective that differentiates our metric from related work.

%\subsubsection{Resilience Measurement Frameworks.}
The measurement framework closest to our work is \emph{Xaminer}, presented by Ramanathan et al.~\cite{ramanathan_2024_xaminer}, that analyzes the impact of physical-layer failures on the network-layer of the Internet.
They do not provide a specific resilience definition, but report the percentage of \gls{IP} traffic that fails due to physical cable failures when addressing resilience against outages.
Most prominently, our framework \toolname supports configuring arbitrary communication networks and application deployments within them, exceeding the functionality of \emph{Xaminer}.

%\subsubsection{Case Studies}
There exist related studies of communication networks facing individual challenges, such as wildfires~\cite{anderson_2020_five},
earthquakes~\cite{mayer_2021_resilience},
solar superstorms~\cite{jyothi_2021_solare},
climate change in general~\cite{durairajan_2018_lights},
and electromagnetic pulse attacks~\cite{neumayer_2011_network,neumayer_2011_assessing}.
In addition, there are case studies of the Internet during previous natural disasters, such as Heidemann et al.~\cite{heidemann_2012_preliminary} and Aben~\cite{aben_2012_ripe}, which report on network outages and routing recovery during Hurricane Sandy, respectively, and Liu et al.~\cite{liu_2012_characterizing} analyzing inter-domain rerouting after the Japan Earthquake.

\subsection{Limitations and Future Work}

This paper includes two case studies in \Cref{sec: pub-f eval france,sec: pub-f eval germany}, but their purpose is to support the contributions of our work rather than evaluate the specific simulated scenarios.
With our case studies, we demonstrate which insights our availability, robustness, and resilience metrics can provide.
We publish \toolname as open-source software under the MIT license to facilitate future work and are curious to see the potential of \toolname when it is applied to support the evaluation in future case studies.
For more complex use cases, we designed \toolname to be extensible and implemented all metrics modularly, so adding new metrics is straightforward.
Doing so only requires implementing the metric as a \emph{Python} function based on our intermediary graph representation of the scenario, and adding a \texttt{streamlit} container that displays the measurement on the web app.

In \Cref{sec: pub-f discussion}, we discuss selected data sources and tools to generate synthetic data that can be integrated into \toolname.
However, the landscape of data sources and tools is constantly evolving, as the resilience of critical infrastructure is still a hot topic.
We designed \toolname to flexibly support models of real-world communication networks, applications, and challenge scenarios.
It expects input data as text files, which can be generated from most available data sources and tools.

% ##################################################################################################################################
% ##################################################################################################################################
% ##################################################################################################################################
% ##################################################################################################################################
% ##################################################################################################################################
\section{Conclusions}

This paper contributes to the resilience of critical infrastructures with three main contributions:
First, we introduce definitions and metrics that capture the end-to-end availability, utility, robustness, and resilience of application deployments (RQ1).
Second, we survey data sources for communication networks, application deployments, and challenges, identifying many datasets and tools that allow researchers to model custom scenarios.
Third, we implement these metrics in our resilience measurement framework \toolname that enables users to model application deployments within communication models and simulate challenges to measure their resilience (RQ2).
It supports the configuration of communication networks, applications, and challenges as input, making it compatible with a wide range of data sources.
We conduct two case studies to demonstrate the functionality of \toolname and highlight the applicability of our end-to-end resilience definition.
% Mobile messaging in France with the NetMob dataset and iGDB physical Internet links and simulated central, high-availability, and island-ready \gls{5G} deployments in Germany.
These case studies showed how \toolname can be used to model and evaluate a wide range of scenarios based on public and synthetic data.
We hope that the research community profits from \toolname to measure the resilience of applications in future work.

\section*{Availability}
Upon the acceptance of this paper, we will publish \toolname as open-source software to facilitate future work.
\toolname is easily extensible, allowing future work to integrate additional metrics.
In addition, we will publish the input data used for the case studies to enable the reproduction and verification of our evaluation.

\section*{Acknowledgments}
This work was supported by the Federal Ministry of Education and Research of Germany in the project Open6GHub (grant number: 16KISK014), by the Federal Ministry of Research, Technology and Space of Germany in the project Open6GHub+ (grant number: 16KIS2407), and by the LOEWE initiative (Hesse, Germany) within the emergenCITY center [LOEWE/1/12/519/03/05.001(0016)/72]. Parts of our research were performed using data made available by Orange within the Netmob 2023 Challenge~\cite{martinez_2023_netmob}.

% \begin{credits}
% \subsubsection{\ackname} A bold run-in heading in small font size at the end of the paper is
% used for general acknowledgments, for example: This study was funded
% by X (grant number Y).

% \subsubsection{\discintname}
% It is now necessary to declare any competing interests or to specifically
% state that the authors have no competing interests. Please place the
% statement with a bold run-in heading in small font size beneath the
% (optional) acknowledgments\footnote{If EquinOCS, our proceedings submission
% system, is used, then the disclaimer can be provided directly in the system.},
% for example: The authors have no competing interests to declare that are
% relevant to the content of this article. Or: Author A has received research
% grants from Company W. Author B has received a speaker honorarium from
% Company X and owns stock in Company Y. Author C is a member of committee Z.
% \end{credits}

\printbibliography
\end{document}